\documentclass[twoside,twocolumn,9pt]{article}
\usepackage{extsizes}
\usepackage[super,sort&compress,comma]{natbib} 
\usepackage[left=1.5cm, right=1.5cm, top=1.785cm, bottom=2.0cm]{geometry}
\usepackage{balance}
\usepackage{sectsty}
\usepackage{graphicx} 
\usepackage{lastpage}
\usepackage{float}
\usepackage{fancyhdr}
\usepackage{fnpos}
\usepackage[english]{babel}
\addto{\captionsenglish}{%
  
}
\usepackage{array}
\usepackage[symbol]{footmisc}

\usepackage{charter}
\usepackage[T1]{fontenc}
\usepackage[usenames,dvipsnames]{xcolor}
\usepackage{setspace}
\usepackage[compact]{titlesec}
\usepackage{hyperref}
\usepackage{bm}
\definecolor{cream}{RGB}{222,217,201}
\usepackage{changes}
\usepackage{amsfonts,amsmath,amssymb} 

\def\bea{\begin{eqnarray}}
\def\eea{\end{eqnarray}}
\def\la{\langle}
\def\ra{\rangle}

\newcommand{\eref}[1]{Eq.~(\ref{#1})}
\newcommand{\erefs}[1]{Eqs.~(\ref{#1})}

\begin{document}

\pagestyle{fancy}
\thispagestyle{plain}


\twocolumn[
  \begin{@twocolumnfalse}

{\huge\textbf{Collective Ring Formation in Active Matter}}

\noindent{\Large Debraj Dutta\textit{$^{\star} $}, Urna Basu\textit{$^{\dag} $} } \\

\noindent {\it S. N. Bose National Centre for Basic Sciences, Kolkata 700106, India.}\\ 

\noindent{$^\star$~debraj.dutta@bose.res.in.}\\
\noindent{$^\dag$~urna@bose.res.in} \\

{\bf Abstract}\\
\noindent We study the formation of ring-like structures in interacting active particle systems in two dimensions. The emergent structure shows signatures of both spatial and orientational organization. The spatial organization is characterized by the radial distance of a tagged particle from the centroid of the assembly, while orientational organization is characterized by the radial alignment of its self-propulsion direction. We derive exact analytical expressions for the radial and polarization distributions for systems of active Brownian particles  and run-and-tumble particles. While both models exhibit annular steady states, we show that their spatial and orientational organization differ qualitatively in the strongly active regime. A direct comparison of the two models reveals how the nature of the propulsion mechanism leads to the distinction in both the structure of the annulus and the statistics of particle orientations. Our results provide a unified analytical framework for characterizing emergent annular states in active matter and identify robust signatures that distinguish persistent active dynamics with continuous and discrete reorientation.

\end{@twocolumnfalse} \vspace{0.6cm}
]


\section{Introduction}

Active matter comprises self-driven particles that continuously consume energy from their surroundings and convert it into persistent motion~\cite{marchetti2013hydrodynamics, bechinger2016active, caprini2022parental}. Examples of active system range from bacterial colonies~\cite{zhang2010collective, peruani2012collective}, motile cells~\cite{trepat2009physical, lecuit2007cell} and cytoskeletal filaments~\cite{sanchez2012spontaneous, keber2014topology} to synthetic realisations such as catalytic Janus colloids~\cite{howse2007self, palacci2013living} and vibrated granular particles, \cite{deseigne2010collective, narayan2007long}. The continuous injection of energy at the microscopic scale drives these systems far from equilibrium, leading to a wealth of phenomena that have no equilibrium counterpart. The unusual features of active matter systems have stimulated extensive theoretical, numerical and experimental research over the past decades.

One of the hallmarks of active matter systems is the emergence of diverse collective phases resulting from the interplay between self-propulsion and inter-particle interactions. Depending on the nature of activity and interactions, active systems can exhibit  flocking~\cite{vicsek1995novel, gregoire2004onset, chate2008collective, buhl2006disorder}, clustering~\cite{peruani2006nonequilibrium, theurkauff2012dynamic, buttinoni2013dynamical, redner2013structure}, motility-induced phase separation~\cite{tailleur2008RTP, cates2015motility, fily2012athermal, redner2013structure, omar2021phase}, emergence of long-range orientational order~\cite{toner1995long, toner1998flocks, toner2005hydrodynamics, chardac2021topology}, formation of living crystals~\cite{palacci2013living, ginot2015nonequilibrium, bricard2013emergence}, vortex states~\cite{wioland2016directed, wensink2012meso, grossmann2014vortex, sanchez2012spontaneous} and various kinds of pattern formation~\cite{solon2015pattern, chate2020dry, xu2024self, liebchen2017collective}. Remarkably, activity can stabilize spatial organizations  that are impossible in passive systems, giving rise to fundamentally new nonequilibrium states of matter. Understanding the physical mechanisms governing these emergent structures and identifying minimal models that admit analytical treatment remain central challenges in active matter research.

Recent studies have shown that isotropic interactions alone, when combined with persistent self-propulsion, can generate collective spatial and orientational order even in the absence of explicit alignment interactions \cite{caprini2023flocking,dutta2026dispersion, caprini2020time}. This has opened up a new class of minimal models for studying emergent collective behaviour in active systems. Despite these advances, the stationary properties of such ordered states remain poorly understood from an analytical perspective, with most existing studies relying predominantly on numerical simulations. In this work, we investigate one such minimal system consisting of active particles coupled through pairwise attractive interactions, which forms a stationary ring-like structure. We consider two of the most widely studied models of active motion---Run-and-Tumble particles (RTP)~\cite{tailleur2008RTP,malakar2018RTP} and active Brownian particles (ABP)~\cite{basu2018active}---and develop an analytical framework to characterize this collective stationary state. In particular, we derive explicit expressions for the radial distribution of a tagged particle measured from the centre of mass and show that, despite the similar ring-like organisation exhibited by both the systems, the radial distribution displays qualitatively distinct behaviour for RTPs and ABPs in the strongly active regime. We further characterise the radial orientational order through fluctuations of the radial polarization, providing complementary signatures that distinguish between these two classes of dynamics.

The paper is organized as follows. In the next section we define the model and present a brief summary of the main results. In Sec.~\ref{sec:Pl} we analytically characterize the spatial organization by computing the radial distribution for RTP and ABP dynamics separately. 
Section~\ref{sec:Ppsi} is devoted to the study of orientation organization in both cases. A comparative discussion of the spatial and orientational organization in RTPs and ABPs, highlighting the influence of the underlying propulsion dynamics, is presented in Sec.~\ref{sec:comp}. We conclude with some general comments in Sec.~\ref{sec:conclusion}. 

\begin{figure}
    \centering
    \includegraphics[width=8.5cm]{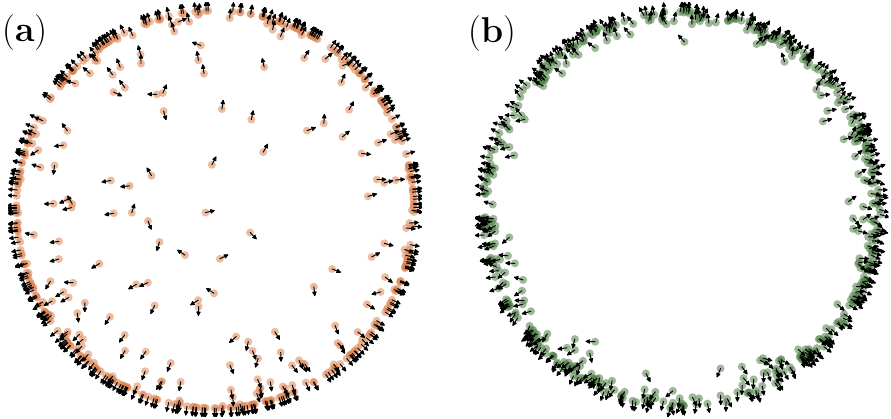}
    \caption{Typical configurations for a collection of $N=500$ harmonically coupled (a) run-and-tumble particles and (b) active Brownian particles. The black arrows indicate the internal orientations of the particles. Here we have taken $\tau=5$, $k=0.002$ and $v_0=1$.}
    \label{fig:cloud}
\end{figure}

\section{Model and Main Results}
We consider a system of \(N\) overdamped active particles moving in two dimensions. In the absence of interactions and external forces, each particle self-propels along an internal orientation that evolves stochastically. In addition, particles interact pairwise through an attractive harmonic potential \(V(r)= \kappa r^2/2\), which depends only on the interparticle separation \(r\). The dynamics of the position \({\mathbf r}_j=(x_j,y_j)\) of the \(j\)-th particle is governed by the Langevin equation
\begin{align}
\dot{\mathbf r}_j
=-\sum_{k\neq j}^{N}\kappa \left(\mathbf r_j-\mathbf r_k\right)+v_0 \hat{\mathbf n}_j, \label{eq:langevin}
\end{align}
where \(\hat{\mathbf n}_j=(\cos\theta_j,\sin\theta_j)\) denotes the orientation vector of the $j$-th particle and \(v_0\) is the self-propulsion speed, assumed to be same for all the particles.

We focus on two different models of self-propulsion which correspond to different evolution dynamics of the orientation vector $\hat{\mathbf n}_j$. First, we consider active Brownian particles~\cite{kurzthaler2018probing,basu2018active} which models motion of self-propelled colloids, swimming microorganisms, and motile cells whose orientation evolves continuously due to rotational diffusion~\cite{howse2007self,jiang2010active}. In this case, the orientation angle of the $j$-th particle undergoes a rotational diffusion,
\begin{align}
\dot{\theta}_j(t)=\sqrt{\frac{2}{\tau}}~\eta_j(t),\label{eq:tht_evol}
\end{align}
where $\{\eta_j(t)\}$ are independent Gaussian white noises satisfying
\begin{align}
\langle \eta_j(t)\eta_k(t')\rangle=\delta_{jk}\delta(t-t'),
\end{align}
and $1/\tau$ is the rotational diffusion coefficient, which is same for all the particles. Next, we consider Run-and-Tumble Particles~\cite{tailleur2008RTP, malakar2018RTP}, which models abrupt changes in orientation vector exhibited by many motile bacteria, such as Escherichia coli and Salmonella enterica~\cite{bergbook}. 
In this case, the orientation $\theta_j$ randomly changes to a new value $\theta_j'$, chosen uniformly  from $[0,2\pi]$, with a rate $1/\tau$.  For both these dynamics, the auto-correlation of the orientation vector decays exponentially,
\begin{align}
   \la \hat {\bm n}_j(t_1) \cdot  \hat{\bm n}_j(t_2) \ra = \exp{[- |t_1 - t_2|/\tau]}. \label{eq:n_auto}
\end{align}
where $\tau$ denotes the persistence time. 

The goal of this work is to analytically characterize the emergent behaviour of this system when the number of particles $N$ is large, and how it depends on the microscopic propulsion mechanism. To this end, it is useful to consider the motion of the particles with respect to the centroid ${\bm R}=\frac{1}{N}\sum_{j=1}^{N}{\bm r}_{j}$. In the centroid coordinate, the Langevin equation governing the time-evolution of the position of the $j$-th particle ${{\bm \ell}_{j}}={\bm r}_j-{\bm R}$ can be expressed as,
\begin{align}
    \dot{\bm \ell}_j=-\mu {\bm \ell}_{j}+v_0b\,\hat{\bm n}_j+ {\bm S}_j,\label{eq:Le_COM}
\end{align}
where,  ${\bm S}_j$ denotes the effective noise acting on the $j$-th particle, and is given by,
\begin{align}
    {\bm S}_j = \frac{v_0}{N}\sum_{k\neq j}\hat{\bm n}_{k},\label{eq:OU_noise}
\end{align}
and, we have defined $\mu=N\kappa$ and $b=(1-1/N)$. It should be noted that the motion of the particles are correlated through the effective noise ${\bm S}_j$, which are not independent. 

\begin{figure}
    \centering
    \includegraphics[width=8.9cm]{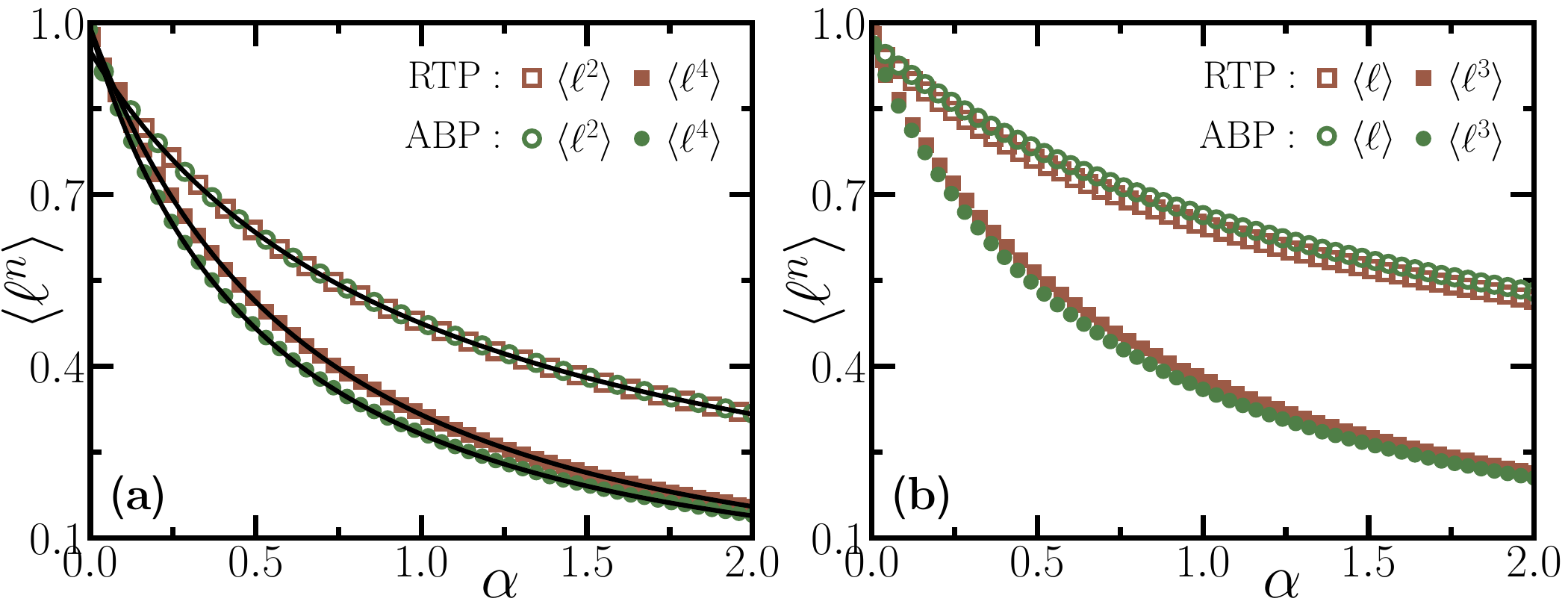}
    \caption{Radial moments: Plot of (a) $\la \ell^2 \ra$, $\la \ell^4 \ra$ and (b) $\la \ell \ra$,$\la \ell^3 \ra$ as functions of activity strength $\alpha=1/(\mu \tau)$ for ABP and RTP dynamics. The symbols indicate data obtained from numerical simulations.  The solid black lines in panel (a) correspond to analytical predictions \erefs{eq:l24_def}-\eqref{eq:l4_avg_ABP}. Here we have used $N=20, k=0.05$ and $v_0=1$.}
    \label{fig:l_moments}
\end{figure}

From \eref{eq:Le_COM} it is clear that the time-evolution of the particle position ${\bm \ell}_j$ is controlled by two time-scales, namely, $\tau$ and $\mu^{-1}$, which denote the persistence time of the self-propulsion direction and relaxation time corresponding to the attractive interaction, respectively. The relative coordinates $\{ \ell_j\}$ reach a stationary state at times much larger than both the time-scales, while the centroid undergoes an unbounded motion.

The interplay of activity and interaction time-scales leads to the emergence of two broad organizational phases in the strongly active regime $\tau \gg \mu^{-1}$ and the passive regime $\tau \ll \mu^{-1}$. In the passive regime, the particles are expected to remain localized near the centroid with typically Gaussian fluctuations. A very different scenario emerges in the active regime, where the particles self-organize into a ring-like spatial structure in the stationary state. This is illustrated in 
Fig.~\ref{fig:cloud} for collections of ABP and RTP. It is apparent from the figure that although the ring-like structure is formed in both cases, the spatial organization differs qualitatively between the two models: for the same set of parameters, the ring formed by the RTPs is more diffuse, with particles more frequently escaping the annulus and wandering toward the centroid than ABPs. From Fig.~\ref{fig:cloud}, it also appears that the system additionally exhibits an orientational order---the internal orientations of the particles on the perimeter of the circle are mostly aligned radially outward with respect to the centroid.  One of the primary focuses of this work is to quantitatively characterize the difference between the two kinds of propulsion mechanisms in the context of the spatial and orientational organisations. 

The spatial organization can be characterized by the radius of the emergent annular structure, which is measured by $\ell_j=|{\bm \ell}_j|$, the distance of a tagged particle from the centroid. On the other hand,  the orientational ordering can be described by the radial polarization $\psi_j =\hat {\bm \ell}_j \cdot \hat{\bm n}_j $, which measures the alignment between the orientation vector and unit position vector $\hat{\bm \ell}_j={\bm \ell}_j/\ell_j$ of the tagged particle. We investigate the properties of these quantities for the ABP and RTP dynamics by computing the corresponding stationary distributions. Before going to the details of the computation we present a brief summary of our main findings here.

\begin{figure}
    \centering
    \includegraphics[width=8.9cm]{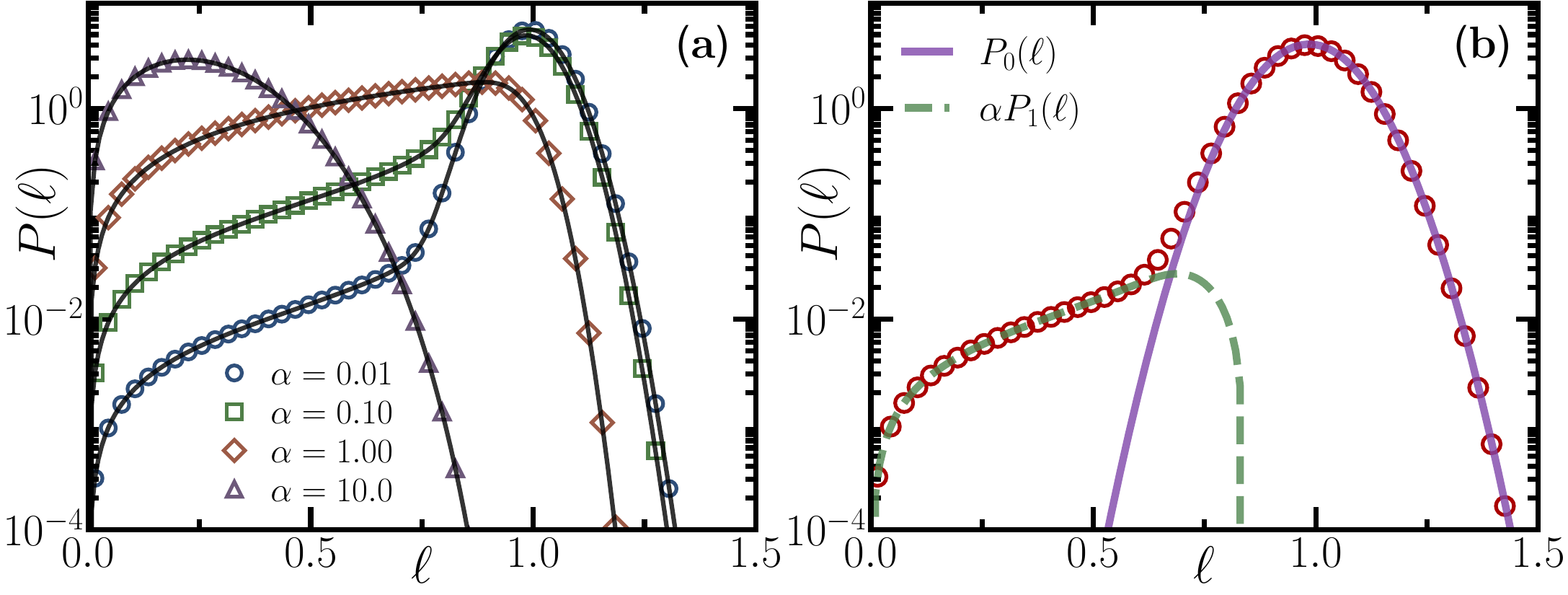}
    \caption{Radial distribution $P(\ell)$ for run-and-tumble dynamics: (a) Plot of $P(\ell)$ for different values of activity strength $\alpha=1/(\mu\tau)$. The symbols indicate data obtained from numerical simulations whereas the solid black lines correspond to the analytical prediction \eref{eq:Pl_RTP}. Here $N=100,k=0.01$ and $v_0=1$, such that $R_0=0.99$. (b) 
    Contributions of $P_0(\ell)$ and $P_1(\ell)$ to 
    the radial distribution $P(\ell)$ for $\alpha=0.01$ i.e., in the strongly active regime. The symbols indicate $P(\ell)$ obtained from numerical simulations whereas the solid and dashed lines correspond to the analytical expressions  \eqref{eq:P0_active} and \eqref{eq:P1_RTP}, respectively. Here we have used $N=50, k=0.02, \tau=100$ and $v_0=1$, so that $R_0=0.98, \sigma\simeq0.0194$.}
    \label{fig:Pl_RTP}
\end{figure}

\begin{itemize}

\item We show that for large number of particles $N$, the emergent noise ${\bm S}_j$ can be effectively described by an Ornstein-Uhlenbeck (OU) process. Using this effective description we obtain the characteristic function $\tilde{P}(q)$ of the stationary distribution $\mathbb{P}(z=\ell^2)$, where $\ell$ denotes the distance of a tagged particle from the centroid. This is formally expressed in terms of the single particle position distribution of the underlying active process [see \eref{eq:Pq_1}].  

\item  Using the  characteristic function, we compute the stationary distribution $P(\ell)$ for both RTP and ABP dynamics [see \erefs{eq:PlRTP_series} and \eqref{eq:Pl_ABP}]. For large $\ell$ both the distributions show similar Gaussian decay, while for small values of $\ell$ these distributions show distinctly different behaviour [see Figs.~\ref{fig:Pl_RTP}, \ref{fig:Pl_ABP} and \ref{fig:compare}]. 

\item We characterise the orientational organisation through the radial polarization $\psi=\hat {\bm \ell} \cdot \hat{\bm n}$ of a tagged particle. Although both mean and variance of $\psi$ decrease with $\alpha$, they show quantitatively distinct behaviour for RTP and ABP dynamics [see Fig.~\ref{fig:psi_moments}]. 

\item We develop a formalism to compute the distribution of the radial polarisation  [see \eref{eq:PPsi_formal}], which is used to obtain the $P(\psi)$ for both RTP and ABP dynamics. It turns out that in the strongly active regime, $P(\psi)$ shows very different qualitative features for RTP and ABP---while $P(\psi)$ shows a single narrow peak near $\psi=1$ for ABP, another peak at $\psi=-1$ emerges for RTP.

\end{itemize}

\section{Spatial Organization}\label{sec:Pl}

In this section, we analytically characterize the emergent spatial organization and how it depends on the self-propulsion mechanism. As mentioned already, a suitable measure of the spatial organisation is given by the distance of a tagged particle from the centroid, which evolves according to \eref{eq:Le_COM}. In the rest of the article, we will discuss the tagged particle behaviour and drop the particle index $j$ for notational simplicity.

To get a preliminary idea of the spatial organisation, we first measure a few moments of the radial distance $\ell$ for both ABP and RTP. This is illustrated in Fig.~\ref{fig:l_moments} which shows that while the second moment $\la \ell^2\ra$ is same for both the systems, $\la \ell \ra$ and higher order moments show significant deviations in the moderate activity regime. In order to further understand the spatial organisation in these two systems, we investigate the probability distribution $P(\ell)$ of the radial distance $\ell$ in the following.

For large number of particles $N\gg 1$, the effective noise ${\bm S}$ defined in \eref{eq:OU_noise} is expected to follow a Gaussian distribution, following central limit theorem. To characterise this Gaussian distribution we first note that $\la{\bm S}\ra=0$ and the autocorrelation of ${\bm S}$ is given by [see \eref{eq:n_auto}],
\begin{align}
    \la S^{(\alpha)}(t)S^{(\beta)}(t')\ra=\delta_{\alpha\beta}\frac{b v_0^2}{2 N}e^{-|t-t'|/\tau}. 
\end{align}
Thus, we can approximate the time evolution of the effective noise $\bm{S}$ using an OU process,
\bea 
\dot{\bm S}=-\frac{\bm S}{\tau}+v_0\sqrt{\frac{b}{N\tau}}{\bm \eta}_S, \label{eq:Le_S}
\eea 
where ${\bm \eta}_S=(\eta_S^{(x)},\eta_S^{(y)})$ is a standard white noise with autocorrelation $\la \eta_{S}^{(\alpha)}(t)\eta_{S}^{(\beta)}(t')\ra=\delta_{\alpha \beta}\delta(t-t')$.

To proceed further, it is convenient to decompose the tagged particle position ${\bm \ell}$ as a sum of two components,
\begin{align}
    {\bm \ell}={\bm \ell}_a+{\bm \ell}_g. \label{eq:l_la_lg}
\end{align}
Equation \eqref{eq:Le_COM} is thus recast in terms of two independent Langevin equations,
\begin{align}
    \dot{\bm \ell}_a &= -\mu{\bm \ell}_a+bv_0\hat{\bm n},\label{eq:Le_active}\\
    \dot{{\bm \ell}}_g &= -\mu{\bm \ell}_g+{\bm S}.\label{eq:Le_OU}
\end{align}
Clearly, ${\bm \ell}_a$ describes the dynamics of a two-dimensional RTP or ABP with self propulsion speed $bv_0$ in a harmonic trap of strength $\mu$. The corresponding stationary radial distribution ${\cal P}(\ell_a)$ is exactly known for both ABP~\cite{malakar2020steady, caraglio2022analytic} and RTP~\cite{Guneau2026}. On the other hand, ${\bm \ell}_g$ denotes an active Ornstein-Uhlenbeck process (AOUP) in the same harmonic trap, which has a Gaussian stationary distribution~\cite{martin2021statistical, bonilla2019active},
\begin{align}
    G(\ell_g) =\frac{2\ell_g}{\sigma^2}e^{-\ell^2_g/\sigma^2},~~~\text{with},~~~\sigma^2=\frac{b v_0^2}{\mu N(\mu+1/\tau)}. \label{eq:G_OU}
\end{align}
The stationary distribution $P(\ell)$ can be computed from \eref{eq:l_la_lg} using the known expressions of ${\cal P}(\ell_a)$ and $G(\ell_g)$ and the fact that $\hat{\bm n}$ and ${\bm S}$ are independent. This can be conveniently done by first determining the distribution  $\mathbb{P}(z)$ of $z={\ell}^2$ and subsequently using the relation,
\begin{align}
    P({\ell})= 2 \ell \mathbb{P}(z=\ell^2).\label{eq:Pzl}
\end{align}
Formally, from \eref{eq:l_la_lg} the stationary distribution $\mathbb{P}(z)$ can be expressed as,
\begin{align}
    \mathbb{P}(z)=\intop_{0}^{\infty}d\ell_a\intop_{0}^{\infty}d\ell_g\intop_{0}^{2\pi}\frac{d\phi}{2\pi}\mathcal{P}(\ell_{a})G(\ell_{g})\delta(z-\ell_a^2-\ell_g^2-2\ell_a\ell_g\cos\phi),\label{eq:dist_l}
\end{align}
where $\phi$ denotes the angle between the vectors ${\bm \ell}_a$ and ${\bm \ell}_g$. It is convenient to first compute the characteristic function of $\mathbb{P}(z)$, 
\begin{align}
    \tilde{P}(q) \equiv \la e^{iqz}\ra = \frac{i}{i+q\sigma^2}\intop_0^{\infty}d\ell_a {\cal P}(\ell_a) e^{-q\ell_a^2/(i+q\sigma^2)}.\label{eq:Pq_1}
\end{align}
To derive the second equation we have performed the integrals over $\phi$ and $\ell_g$ in \eref{eq:dist_l} using \eref{eq:G_OU}.

To proceed further, we need to specify the underlying active dynamics which determines ${\cal P}(\ell_a)$. In the following, we consider the two most well known cases of active particle dynamics, namely, RTP and ABP, separately.

\subsection{Run-and-Tumble particles}\label{sec:Pl_RTP}
We start by considering the case of run-and-tumble dynamics. The stationary radial distribution of an RTP following \eref{eq:Le_active} is given by~\cite{Guneau2026},
\begin{align}
    {\cal P}(\ell_a)=\frac{2\alpha\ell_a}{R_0^2}\Bigg[1-\Big(\frac{\ell_a}{R_0}\Big)^2\Bigg]^{\alpha-1}\Theta(R_0-\ell_a),\label{eq:P_RTP}
\end{align}
where $\alpha=1/(\tau\mu)$ and $R_0=bv_0/\mu$. The dimensionless parameter $\alpha$ provides a measure of the activity of the particle. A small value of $\alpha \ll 1$, corresponding to a large persistence time $\tau$, indicates that the system is strongly active, while large values of $\alpha$ indicate a passive regime. In the strongly active regime ($\alpha \ll 1$) the single particle distribution \eref{eq:P_RTP} diverges near $\ell_a=R_0$. The shape of the distribution changes as $\alpha$ in increased, eventually resembling a Rayleigh-like distribution in the strongly passive regime.

Substituting the above equation in \eref{eq:Pq_1} and integrating over $\ell_a$, we obtain an explicit expression for the characteristic function \eqref{eq:Pq_1} in the interacting system,
\begin{align}
    \tilde{P}(q)=\frac{i}{i+q\sigma^2}~_1F_1(1,1+\alpha,-\frac{R_0^2q}{i+q\sigma^2}),\label{eq:RTP_char}
\end{align}
where $_1F_1(a,b,z)$ is the confluent hypergeometric function of the first kind \cite[\href{https://dlmf.nist.gov/13.2.E3}{(13.2.3)}]{NIST:DLMF}. 
Note that $R_0$ here is the characteristic length scale associated with the RTP dynamics, while $\sigma$ is the characteristic length scale associated with the Gaussian AOUP dynamics. Together with the activity strength $\alpha$, the interplay of these length scales determine the nature of the steady-state distribution.

To obtain $\mathbb{P}(z)$ explicitly, we need to invert the Fourier transform, which can be performed using an integral representation of the confluent hypergeometric function which leads to [see Appendix~\ref{ap:Pl_RTP} for details], 
\begin{align}
    P(\ell) = \frac{2\alpha \ell}{\sigma^2}e^{-\ell^2/\sigma^2} \intop_0^1dt\,(1-t)^{\alpha-1}e^{-R_0^2 t/\sigma^2}I_0\left(\frac{2R_0\ell}{\sigma^2}\sqrt{t}\right),\label{eq:Pl_RTP}
\end{align}
where, $I_0(z)$ is the zeroth order modified Bessel function of the first kind~\cite[\href{https://dlmf.nist.gov/10.25.2}{(10.25.2)}]{NIST:DLMF} Note that we have used \eref{eq:Pzl} to obtain $P(\ell)$ from $\mathbb{P}(z)$. For any values of the parameters $v_0, \tau, k, $ and $N$, the stationary distribution $P(\ell)$ can be estimated to arbitrary accuracy by performing the integral numerically. However, we can get a more explicit expression using the series reperesentation of $I_{0}(x)$ and performing the integral over $t$, 
\begin{align}
    P(\ell) = \frac{2\ell}{\sigma^2}e^{-\frac{\ell^2}{\sigma^2}}\sum_{k=0}^{\infty}\frac {\alpha\Gamma(\alpha)}{k!(k+\alpha)!}\Big(\frac{R_0\ell}{\sigma^2}\Big)^{2k}{}_1{F}_{1}\left[k+1,\alpha+k+1,-\frac{R_0^2}{\sigma^2}\right].\label{eq:PlRTP_series}
\end{align}
Figure~\ref{fig:Pl_RTP}(a) shows a plot of the radial distribution $P(\ell)$ obtained from numerical simulations along with the theoretical prediction \eref{eq:Pl_RTP} for different values of $\alpha$. The excellent match between the two shows that the effective \erefs{eq:Le_COM} and \eqref{eq:Le_S} describe the system accurately in the large $N$ limit. 

\begin{figure}[t]
    \centering
    \includegraphics[width=7.0cm]{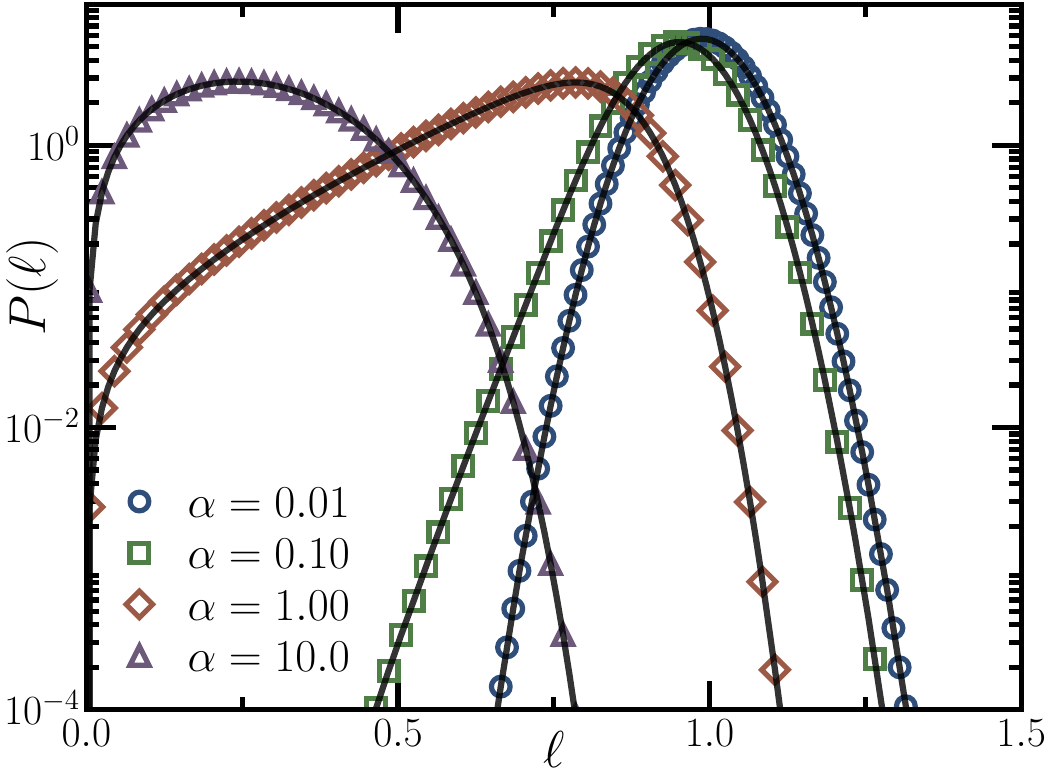}
    \caption{Radial distribution $P(\ell)$ for active Brownian particles: Plot of $P(\ell)$ for different values of activity strength $\alpha=1/(\mu\tau)$. The symbols indicate data obtained from numerical simulations whereas the solid black lines correspond to the analytical prediction \eref{eq:Pl_ABP}. Here we have used $N=100,k=0.01$ and $v_0=1$, such that $R_0=0.99$.}
    \label{fig:Pl_ABP}
\end{figure}

From Fig.~\ref{fig:Pl_RTP}(a), it appears that for small $\alpha\ll1$, i.e., in the strongly active regime, the distribution shows an interesting shape---a Gaussian-like peak around $\ell=R_0$, accompanied by a shoulder like structure in the small $\ell$ regime. This shoulder disappears as $\alpha$ is increased leading to a broad distribution which eventually crosses over to a Gaussian-like one in the passive regime $\alpha\gg1$.
To understand this shape transition of $P(\ell)$, we next look at the active and passive regimes separately.  

\subsubsection{Active regime} We first look at the strongly active regime $\alpha \ll 1$. Physically, we expect the distribution $P(\ell)$ to be a well behaved function of $\alpha$. Thus, in this regime, we can write 
\begin{align}
    P(\ell) = P_0(\ell)+\alpha P_1(\ell)+O(\alpha^2).\label{eq:Pl_ac_approx}
\end{align}
Here, $P_0(\ell)$ corresponds to the distribution in the $\alpha\to0$ limit, i.e., when the persistence time $\tau\to\infty$. From \eref{eq:Pl_RTP}, we have, 
\begin{align}
    P_0(\ell) = \frac{2\ell}{\sigma^2}e^{-(\ell^2+R_0^2)/\sigma^{2}}I_0\Big(\frac{2R_0\ell}{\sigma^2}\Big).\label{eq:P0_active}
\end{align}
For large $\ell$, this shows a Gaussian like decay with a peak of width $\sigma\sim1/N^{3/2}$, 
\begin{align}
    P_0(\ell)\approx\sqrt{\frac{\ell}{\pi \sigma^2R_0}} e^{-(\ell-R_0)^2/\sigma^2}.\label{eq:large_l_active}
\end{align}
On the other hand for small $\ell$, we have,
\begin{align}
    P_0(\ell) \approx\frac{2\ell}{\sigma^2}e^{-R_0^2/\sigma^2}\approx\frac{2\ell}{\sigma^2}e^{-N}.\label{eq:P0_sml_RTP}
\end{align}
Clearly, the weight near $\ell=0$ vanishes exponentially with the particle number $N$. Thus $P_0(\ell)$ corresponds to the Gaussian-like peak around $\ell=R_0$ [see Fig.~\ref{fig:Pl_RTP}]. 

For finite but large $\tau$, i.e., in the regime $0<\alpha\ll 1$, the leading order correction can be obtained by expanding the confluent hypergeometric function around $\alpha=0$ [see Appendix~\ref{ap:Pl_RTP} for details] and we have, 
\begin{align}
    P_1(\ell) &\simeq \frac{2\ell}{\sigma^2}e^{-\ell^2/\sigma^{2}}\Big[g(\ell)-E_\gamma e^{-R_0^2/\sigma^2}I_0\Big(\frac{2R_0\ell}{\sigma^2}\Big)\Big].\label{eq:P1_RTP}
\end{align}
Here $E_\gamma=0.5772\ldots$ denotes the Euler gamma constant and we have defined,
\begin{align}
   g(\ell) = \sum_{k=0}^{\infty} \frac{1}{k!}\Big(\frac{R_0\ell}{\sigma^2}\Big)^{2k}\Bigg[\partial_\alpha\frac{{}_1{F}_{1}(k+1,k+1+\alpha,-\frac{R_0^2}{\sigma^2})}{(k+\alpha)!}\Bigg]_{\alpha=0}.
\end{align}
Interestingly, we have, for small $\ell\ll R_0$ [see Appendix~\ref{ap:Pl_RTP}],
\begin{align}
    P_1(\ell)\approx\frac{2\ell}{\sigma^2}\frac{\sigma^2}{ R_0^2}\approx\frac{2\ell}{\sigma^2}\frac{1}{N}.\label{eq:P1_sml_RTP}
\end{align}
Comparing \erefs{eq:P0_sml_RTP} and \eqref{eq:P1_sml_RTP}, it is clear that, 
$P_1(\ell)$ contributes significantly to the behaviour of the distribution in the small $\ell$ regime, giving rise to the shoulder observed in Fig.~\ref{fig:Pl_RTP}(a). 
This is illustrated in Fig.~\ref{fig:Pl_RTP}(b) which shows the contributions of $P_0(\ell)$ and $P_1(\ell)$ separately.

\subsubsection{Passive Regime}

The strongly passive regime emerges when the 
persistence time $\tau$ is much smaller than the interaction time-scale $\mu^{-1}$. Mathematically, in the $\tau \to 0$ limit, the auto-correlation of $\hat{\bm n}(t)$ in \eref{eq:n_auto} reduces to a delta-function~\cite{santra2021DRABP}. Thus, the typical fluctuations of $\ell$ can be understood by replacing ${\bm \hat n(t)}$ with a white noise of strength $b^2v_0^2\tau$. Moreover, in this limit, ${\bm S}$ also emulates an independent white noise. Hence, \eref{eq:Le_COM} implies that the typical fluctuations ${\bm \ell}$ are similar to that of a 2D Brownian motion in a harmonic trap. Correspondingly, the typical fluctuations of the  radial component $\ell$ are expected to be described by a Rayleigh distribution,
\begin{align}
    P(\ell)\approx\frac{2\ell}{\sigma^2+R_0^2/\alpha} \exp\Big[-\frac{\ell^2}{\sigma^2+R_0^2/\alpha}\Big]. \label{eq:rayleigh}
\end{align}
This is also illustrated in Fig.~\ref{fig:Pl_RTP} where the dashed curve corresponds to \eref{eq:rayleigh} which shows excellent agreement with the $P(\ell)$ obtained from numerical simulations for large $\alpha$ value. The above result can also be derived more rigorously from \eref{eq:Pl_RTP} by considering the large $\alpha$ regime [see Appendix~\ref{ap:Pl_RTP} for details].

\subsection{Active Brownian particles}\label{sec:Pl_ABP}

In section we consider the case of ABP, i.e., when the orientation of the particles evolve following \eref{eq:tht_evol}. The position distribution of an ABP in harmonic confinement is explicitly known only in the presence of a translational thermal noise and is given as an infinite series sum~\cite{malakar2020steady}.
Translating this solution in terms of our parameters, we get,
\begin{align}
    {\cal P}(\ell_a) = \frac{\ell_a}{D_t}\sqrt{\frac{2\pi \mu}{\tau}}e^{-\mu \ell_a^2/(2D_t)}\sum_{n=0}^{\infty}\Big(\frac{b^2v_0^2\tau}{D_t}\Big)^n C_{n,0}\, L_n\Big(\frac{\mu\ell_a^2}{2D_t}\Big),\label{eq:P_ABP}
\end{align}
where $L_{n}(x)$ is the $n$-th order Laguerre polynomial and $D_t$ denotes the translation diffusion coefficient. The coefficients $C_{n,0}$ depend on $\alpha=1/(\mu\tau)$ and satisfy the recursion relation\cite{malakar2020steady},
\begin{subequations}
\begin{align}
    C_{n,l}&=\frac{C_{n,l-1}\sqrt{(n+l)/(2\alpha)}-C_{n-1,l+1}\sqrt{n/(2\alpha)}}{(2n+l)/\alpha+l^2},~~l>0,\label{eq:C_recur_1}\\
    C_{n,0}&=-C_{n-1,1}/\sqrt{2n/\alpha},\label{eq:C_recur_2}
\end{align}\label{eq:recur}
\end{subequations} 
with boundary condition $C_{0,0}=\sqrt{\mu\tau/(2\pi)}$. In the strongly active regime, i.e., when $\mu\tau \gg 1$, the distribution is sharply peaked around $R_0 = b v_0/\mu$. As $\tau$ is decreased, it crosses over to a Rayleigh-like distribution expected in the strongly passive regime $\mu\tau \ll 1$.

Note that, our case corresponds to $D_t =0$, where ${\cal P}(\ell_a)$ becomes singular. However, in the following, we show that after performing the convolution with the AOUP distribution, this limit becomes well defined, yielding the exact result required for our case. 

We start from  for the characteristic function $\tilde P(q)$ given in \eref{eq:Pq_1}. Using \eref{eq:P_ABP} in \eref{eq:Pq_1} and performing the integral over $\ell_a$ [see Appendix~\ref{ap:Pl_ABP} for the details of this computation], we arrive at,
\begin{align}
    \tilde{P}(q) = i\sqrt{\frac{2\pi}{\mu\tau}}\sum_{n=0}^{\infty}C_{n,0}\Big(\frac{2\tau b^2v_0^2}{\mu}\Big)^{n}\frac{q^n}{\left[i+q(\sigma^2+2D_t/\mu)\right]^{n+1}},\label{eq:ABP_char}
\end{align}
where, $\sigma^2$ is defined in \eref{eq:G_OU}. At this stage, we can safely take the limit $D_t\to0$ which leaves a well-defined $\tilde P(q)$. To compute $\mathbb{P}(z)$, we need to take inverse Fourier transform of \eref{eq:ABP_char} with respect to $q$,
\begin{align}
    \mathbb{P}(z) = i\sqrt{\frac{2\pi}{\mu\tau}}\sum_{n=0}^{\infty}C_{n,0}\, \Big(\frac{2 \tau b^2v_0^2}{\mu}\Big)^{n}\intop_{-\infty}^{\infty}\frac{dq} {2\pi}\frac{e^{-iqz}q^n}{(i+q\sigma^2)^{n+1}}.\label{eq:Pz_1}
\end{align}
The above integral has an $(n+1)$-th order pole at $q_0=-i/\sigma^2$, and can be evaluated using Cauchy Residue theorem. The details of the calculation are given in Appendix~\ref{ap:Pl_ABP},
here we give the final expression,
\begin{align}
    P(\ell) = \sqrt{2\pi\alpha}\, \Big(\frac{2\ell}{\sigma^2}\Big)e^{-\ell^2/\sigma^2}\sum_{n=0}^{\infty}C_{n,0}\,\Big(\frac{2R_0^2}{\alpha\sigma^2}\Big)^n L_{n}\Big(\frac{\ell^2}{\sigma^2}\Big),\label{eq:Pl_ABP}
\end{align}
where we have introduced the same parameters $\alpha = 1/(\tau\mu)$ and $R_0=bv_0/\mu$ as defined for the RTP case. 

Clearly, the radial distribution for the ABP given by the above equation is very different from that for the RTP dynamics [see \eref{eq:PlRTP_series}]. This is also shown in Fig.~\ref{fig:Pl_ABP} which shows how $P(\ell)$ changes as $\alpha$ is increased. For small values of $\alpha$, i.e., in the strongly active regime, the distribution has a single-peaked Gaussian-like shape. As $\alpha$ is increased, the distribution becomes broader with the peak shifting towards the origin, eventually converging to the Rayleigh distribution expected in the passive regime. It should be emphasized that the shape of $P(\ell)$ in the strongly active regime is markedly different from the RTP case. We provide a detailed comparison between the ABP and RTP cases later in Sec.~\ref{sec:comp}.

In the following we look at the behaviour of the radial distribution in the active and passive regimes separately. 

\begin{figure}
    \centering
    \includegraphics[width=7.0cm]{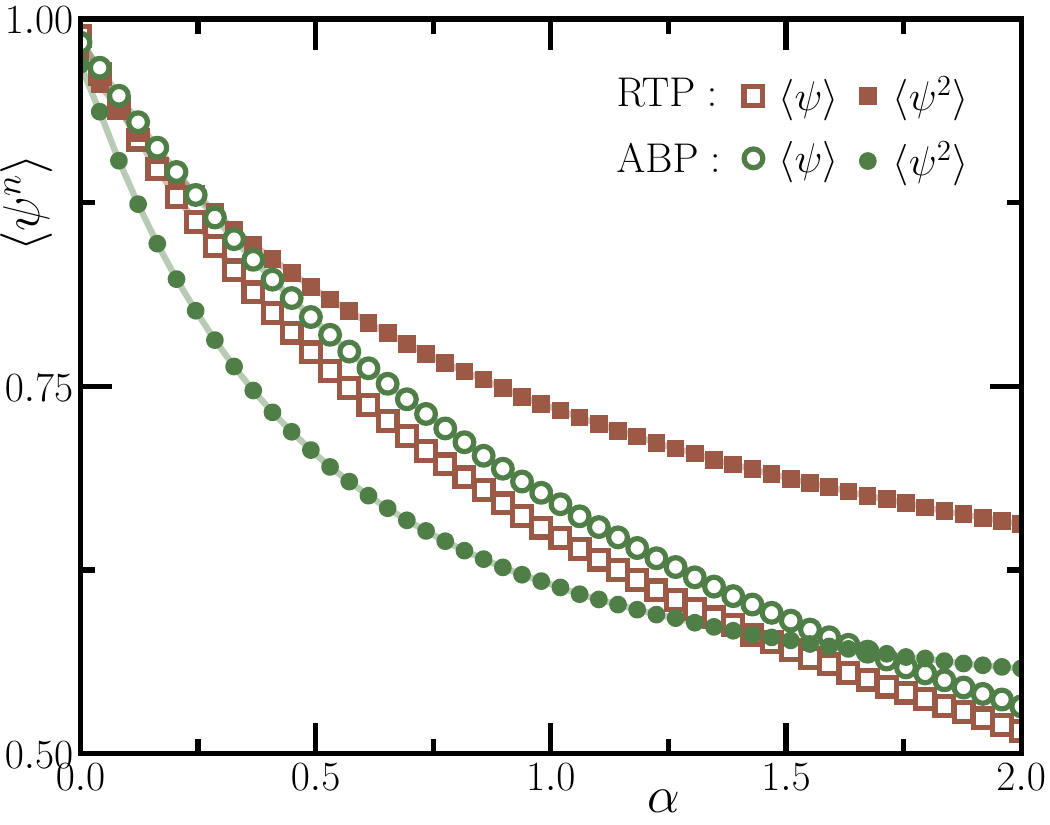}
    \caption{Moments of radial polarization $\psi$: Plots of $\la \psi \ra$ and $\la \psi^2 \ra$ as functions of $\alpha$ for RTP and ABP dynamics. The parameters used are $N=20, k = 0.05$ and $v_0=1$.}
    \label{fig:psi_moments}
\end{figure}

\subsection{Active Regime}

In this section we characterise the radial distribution $P(\ell)$ in the strongly active regime, i.e., when $\alpha\ll1$. To extract the small $\alpha$ behaviour from \eref{eq:Pl_ABP}, we first need to know the $\alpha$ dependence of the coefficients $C_{n,0}$. The explicit forms of these coefficients to leading order in $\alpha$ were obtained earlier~\cite{malakar2020steady} and are given by, 
\begin{align}
    C_{n,l}^{(0)} = \frac{1}{\sqrt{2\pi\alpha}}\frac{(-1)^{n}}{\sqrt{n!(n+l)!(2/\alpha)^{2n+l}}}.\label{eq:C_0}
\end{align}
Substituting \eref{eq:C_0} in \eref{eq:Pl_ABP}, we obtain the $\alpha\to0$ limit of the radial distribution $P_0(\ell)$ for the ABPs, which is identical to the corresponding result for the RTP case given in \eref{eq:P0_active}---a Gaussian-like shape with the peak near $\ell=R_0$. This indicates that the difference between the RTP and ABP cases appear only in higher order terms in $\alpha$. To quantitatively estimate this difference, we need to obtain first the next order correction to the distribution, so that $P(\ell)\approx P_0(\ell)+\alpha P_{1}(\ell)$. To obtain $P_1(\ell)$, we require the higher order corrections in $C_{n,l}$, which can be systematically extracted using the recursion relations \erefs{eq:recur} [see Appendix~\ref{ap:Pl_ABP} for details of the computation],
\begin{align}
    C_{n,l} \approx C_{n,l}^{(0)}\Big(1-\alpha\Big[n+\frac{l(l+1)}{2}\Big]\Big).\label{eq:C_approx_active}
\end{align}
Substituting \eref{eq:C_approx_active} in \eref{eq:Pl_ABP}, and subsequently carrying out the sum gives,
\begin{align}
    P_{1}(\ell)=\frac{2\ell R_0^2}{\sigma^4}e^{-(\ell^2+R_0^2)/\sigma^2}\Big[I_0\Big(\frac{2R_0\ell}{\sigma^2}\Big)-\frac{\ell}{R_0}I_1\Big(\frac{2R_0\ell}{\sigma^2}\Big)\Big],\label{eq:P1_active_ABP}
\end{align}
which is different from the corresponding correction in RTP given in \eref{eq:P1_RTP}. This difference becomes most prominent in the small $\ell$ regime, where,
\begin{align}
    P(\ell)\approx\frac{2\ell R_0^2}{\sigma^4}e^{-R_0^2/\sigma^2}  \approx \frac{2\ell}{\sigma^2}\,N e^{-N},
\end{align}
vanishes as $e^{-bN}$ with particle number $N$, in contrast to the $1/N$ correction for RTP. Consequently the shoulder like structure is absent in the ABP distribution.

\subsection{Passive Regime}
The heuristic argument used to obtain the radial distribution in the passive regime of RTP also holds here. Thus in the regime $\alpha\gg1$, we expect the Rayleigh distribution given in \eref{eq:rayleigh} to describe the typical fluctuations of $\ell$ also for the ABP case. 

The radial distribution in the passive regime can also be obtained more rigorously from \eref{eq:Pl_ABP} by using the large $\alpha$ asymptotic form of the coefficients obtained earlier~\cite{malakar2020steady},
\begin{align}
    C_{n,0}\approx\frac{1}{\sqrt{2\pi\alpha}}\Big(-\frac{1}{2}\Big)^{n}.\label{eq:C_approx_passive}
\end{align}
Substituting \eref{eq:C_approx_passive} in \eref{eq:Pl_ABP} and subsequently evaluating the sum leads to the expected Rayleigh distribution \eqref{eq:rayleigh}. 

For ABP, the next order corrections in $1/\alpha$ to the coefficients $C_{n,0}$ in \eref{eq:C_approx_passive} is also known~\cite{malakar2020steady},
\begin{align}
    C_{n,0}\approx\frac{1}{\sqrt{2\pi\alpha}}\Big(-\frac{1}{2}\Big)^{n}\Big[1-\frac{1}{8\alpha}(7n^2+n)\Big].
\end{align}
Using the above equation, we can compute the correction to the Rayleigh distribution in the passive regime [see Appendix~\ref{ap:Pl_ABP}],
\begin{align}
P(\ell)=\frac{2\alpha \mu^2\ell}{v_0^2}e^{-\alpha\mu^2\ell^2/v_0^2}
\left[1+\frac{5\ell^2\mu^2}{2v_0^2}
-\frac{3}{4\alpha}-\frac{7\alpha}{8}
\left(\frac{\ell\mu}{v_0}\right)^4
\right]. \label{eq:rayleigh_corr}
\end{align}

\begin{figure}
    \centering
    \includegraphics[width=8.9cm]{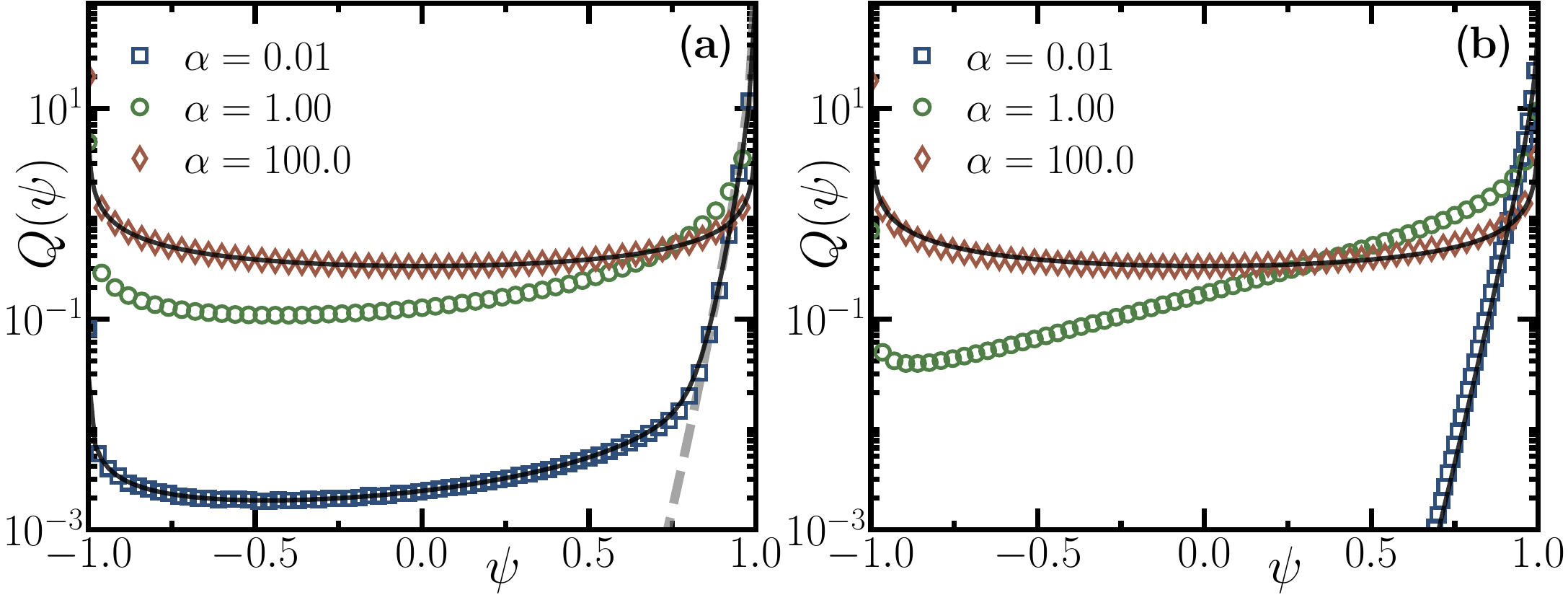}
    \caption{Radial polarization distribution: Plot of $Q(\psi)$ for (a) RTP dynamics and (b) ABP dynamics, for different values of activity $\alpha$. The symbols correspond to data obtained from numerical simulations whereas the black solid lines indicate the corresponding analytical expressions for strongly active [see \eref{eq:Q_F_s} for RTP and \eref{eq:Ppsi_ABP} for ABP] and passive cases [see \eref{eq:Ppsi_passive}]. The dashed line in panel (a) corresponds to the $\alpha \to 0$ limiting result [see \eref{eq:F_0_limit}]
    For both panels we have used $N=20, k=0.05$ and $v_0=1$, so that $R_0=0.95$.}
    \label{fig:Ppsi}
\end{figure}

\section{Orientational Organization} \label{sec:Ppsi}
In this section we analytically characterise the orientational organisation using the radial polarization $\psi={\bm \ell}\cdot\hat{\bm n}$, which is bounded in the regime $[-1, 1]$. Clearly, $\psi\simeq1$ indicates that the orientation of the tagged particle is parallel to the radial vector, while $\psi\simeq-1$ implies anti-parallel alignment between the two. A direct measure of the orientational organisation is thus provided by $\la \psi \ra$, which along with the second moment $\la \psi^2 \ra$ are shown in Fig.~\ref{fig:psi_moments}. Clearly, $\la \psi \ra$ and $\la \psi^2 \ra$ show qualitatively similar behaviour for both ABP and RTP dynamics---both of them monotonically decrease with $\alpha$, indicating reduction in orientational organization as activity is decreased. However, the quantitatively the behaviour for RTP and ABP dynamics is distinctly different. 

To further explore the nature of this radial polarization, we measure the distribution $Q(\psi)$ for both ABP and RTP, which is shown in Fig.~\ref{fig:Ppsi} for different values of $\alpha$. Clearly, the distributions are qualitatively similar in the passive regime, while they have strikingly different shape in the strongly active regime. For RTP  $Q(\psi)$ diverges around $\psi=\pm1$ for all values of $\alpha$, the divergence near $\psi=1$ being stronger for small $\alpha$ [see Fig.~\ref{fig:Ppsi}(a)]. On the other hand, for ABP $Q(\psi)$ vanishes near $\psi=-1$ in the strongly active regime [see Fig.~\ref{fig:Ppsi}(b)]. 

In the following, we analytically compute the radial polarization distribution in the strongly active and passive limits for ABP and RTP separately. Formally the distribution of the radial polarization is given by, 
\begin{align}
    Q(\psi) = \frac{1}{2\pi}\intop_0^{2\pi}d\theta\intop d{\bm \ell}\,P({\bm \ell}|\hat{\bm n})\delta(\psi-\hat{\bm n}\cdot\hat{\bm \ell}),\label{eq:PPsi_formal}
\end{align}
where we have used the fact that in the stationary state, the orientation angle $\theta$ is uniformly distributed in $[0, 2\pi]$. In the passive limit, i.e., when $\alpha\to\infty$, irrespective of the propulsion mechanism, $\hat{\bm n}$ becomes equivalent to a white noise and we have,
\begin{align}
    P({\bm \ell}|\hat{\bm n})=P({\bm \ell}) = \frac{\mu}{\pi D_\text{eff}} e^{-\mu \ell^2/D_\text{eff}},\label{eq:P_ln}
\end{align}
where $D_\text{eff}=\mu\sigma^2+\mu R_0^2/\alpha$. Using \eref{eq:P_ln} in \eref{eq:PPsi_formal} and performing the integral over $\ell$ we have,
\begin{align}
    Q(\psi) = \intop_0^{2\pi}\frac{d\chi}{2\pi} \delta(\psi-\cos\chi)=\frac{1}{\pi\sqrt{1-\psi^2}},\label{eq:Ppsi_passive}
\end{align}
where, $\chi$ in the first equation denotes the angle between ${\bm \ell}$ and $\hat{\bm n}$. From Fig.~\ref{fig:Ppsi}, it is clear that the above distribution accurately describes the behaviour of $Q(\psi)$ in the strongly passive, i.e., the $\alpha\gg1$, regime for both RTP and ABP. 

To understand the behaviour of $Q(\psi)$ in the active regime, we first note that the conditional distribution $P({\bm \ell}|\hat{\bm n})$ can be formally expressed as,
\begin{align}
    P({\bm \ell}| \hat{\bm n}) = \intop d{\bm \ell}_{g}\intop d{\bm \ell}_{a} P({\bm \ell}_a, {\bm \ell}_g|\hat{\bm n}) \delta^{(2)}({\bm \ell} - {\bm \ell}_a - {\bm \ell}_g).\label{eq:ln_joint}
\end{align}
Remembering that ${\bm \ell_g}$ is independent of $({\bm \ell}_a,\hat{\bm n})$ and follows a Gaussian distribution [see \eref{eq:G_OU}], \eref{eq:ln_joint} can be simplified to,
\begin{align}
    P({\bm \ell}| \hat{\bm n}) = \frac{1}{\pi\sigma^2}\intop d{\bm \ell}_{a}P({\bm \ell}_a|\hat{\bm n}) e^{-({\bm \ell} - {\bm \ell_a})^2/\sigma^2}.\label{eq:P_ln_cond}
\end{align}
Next, we compute the conditional distribution $P({\bm \ell}_a|\hat{\bm n})$ and the corersponding $P(\psi)$ separately for RTP and ABP dynamics.

\subsection{Run-and-Tumble particles}\label{sec:Ppsi_RTP}

To compute the conditional distribution $P({\bm \ell}_a|\hat{\bm n})$ for the RTPs, we start with the Langevin equation \eqref{eq:Le_active}. Between consecutive tumbles, the orientation vector $\hat{\bm n}$ remains fixed and \eref{eq:Le_active} can be solved to write,
\begin{align}
    {\bm \ell}_a(t) = {\bm \ell}_a(0) e^{-\mu t}+R_0\hat{\bm n}(1-e^{-\mu t}).\label{eq:l_sol}
\end{align}
In the strongly active regime, the typical value of $\psi$ remains close to unity, and, remembering that 
$P(\ell)$ is strongly peaked around $\ell=R_0$, 
as a first approximation, we write ${\bm \ell}(0)=R_{0}\hat{\bm n}'$. Defining the variable $s=e^{-\mu t}\in[0,1]$, \eref{eq:l_sol} can thus be recast in the form,
\begin{align}
    {\bm \ell}_a(t) = R_0 [(1-s) \hat{\bm n} +s \hat{\bm n}'].
\end{align}
This allows us to write the conditional probability, by summing over contributions from all $\hat{\bm n}'(\phi)\in[0,2\pi]$ and run times through $s\in(0,1)$,
\begin{align}
    P({\bm \ell}_a|\hat{\bm n}) = \intop_0^1 ds \rho(s)\intop_0^{2\pi}\frac{d\phi}{2 \pi} \delta^{(2)}({\bm \ell}_a-R_0\hat{\bm n}(1-s)-R_0\hat{\bm n}'(\phi)s), \label{eq:Pla_n_rtp}
\end{align}
where $\rho(s)$ denotes the probability distribution of $s$. Since the run-times are exponentially distributed with rate $1/\tau$, we have $\rho(s)=\alpha s^{\alpha-1}$. 

Substituting \eref{eq:Pla_n_rtp} in \eref{eq:P_ln_cond} we can formally write,
\bea 
Q(\psi) = \intop_0^1 ds \rho(s) F(s),
\eea 
where, we have defined,
\begin{align}
    F(\psi,s) = \frac 1{\pi \sigma^2}\intop_0^{2 \pi} \frac{d \theta}{2 \pi} \int d{\bm \ell}   \intop_{0}^{2\pi}\frac{d\phi}{2\pi}e^{-\big|{\bm \ell}-R_0[(1-s)\hat{\bm n}) - s\hat{\bm{n}}']\big|^2/\sigma^2} \delta(\psi - \hat{\bm n}\cdot \hat{\bm \ell}). \label{eq:Fs}
\end{align}
Since we are interested in the strongly active regime, where $\alpha \to 0^+$, we can approximate the distribution $\rho(s) \simeq \delta(s) + \alpha /s$ in the integral. Using the subsequent expression in \eref{eq:PPsi_formal} we find, to linear order in $\alpha$, 
\begin{align}
    Q(\psi) = F(\psi,0) + \alpha \intop_0^1 \frac{ds}{s} (F(\psi,s)-F(\psi,0)). \label{eq:Q_F_s}
\end{align}
Physically, the first term corresponds to $\alpha=0$,  i.e., the deterministic scenario where the orientations of the particles remain frozen. The second term gives the first order correction emerging from the evolution of the orientation vector.

The contribution for $\alpha=0$ can be explicitly obtained by performing the integrals in \eref{eq:Fs}, which leads to [see Appendix~\ref{ap:Ppsi_RTP} for the details],
\begin{align}
    F(\psi, 0) = \frac{e^{-R_0^2/\sigma^2}}{\pi\sqrt{1-\psi^2}}\Big[1+\sqrt{\pi}\frac{R_0\psi}{\sigma}e^{\psi^2/\sigma^2}\text{erfc}\left(-\frac{R_0\psi}{\sigma}\right)\Big].\label{eq:PPsi_a0}
\end{align}
For large $N$, $R_0/\sigma\gg1$ and the first term gets exponentially suppressed, $\text{erfc}({-R_0\psi/\sigma})\to2$, yielding,
\begin{align}
    F(\psi,0) \approx  \frac{\sqrt{2}R_0}{\sigma}\frac{\psi}{\sqrt{1-\psi^2}}e^{-R_0^{2}(1-\psi^2)/\sigma^2}.\label{eq:F_0_limit}
\end{align}
This corresponds to the strong peak of the distribution 
near $\psi=1$ [see Fig.~\ref{fig:Ppsi}(a)]. 

The $O(\alpha)$ corrections to $Q(\psi)$ can be obtained by first performing the ${\bm \ell}$ and $\theta$ integrals in \eref{eq:Fs}, which reduces $F(\psi, s)$ to a single integral [see Appendix~\ref{ap:Ppsi_RTP}]. Using the resulting expression along with \eref{eq:Q_F_s} and evaluating the integrals numerically we obtain $Q(\psi)$ to linear order in $\alpha$. This is compared with numerical simulations in Fig.~\ref{fig:Ppsi}, which shows an excellent agreement. Clearly, the $\alpha$-order correction gives rise to another peak near $\psi=-1$. 

\begin{figure}[t]
    \centering
    \includegraphics[width=8.9cm]{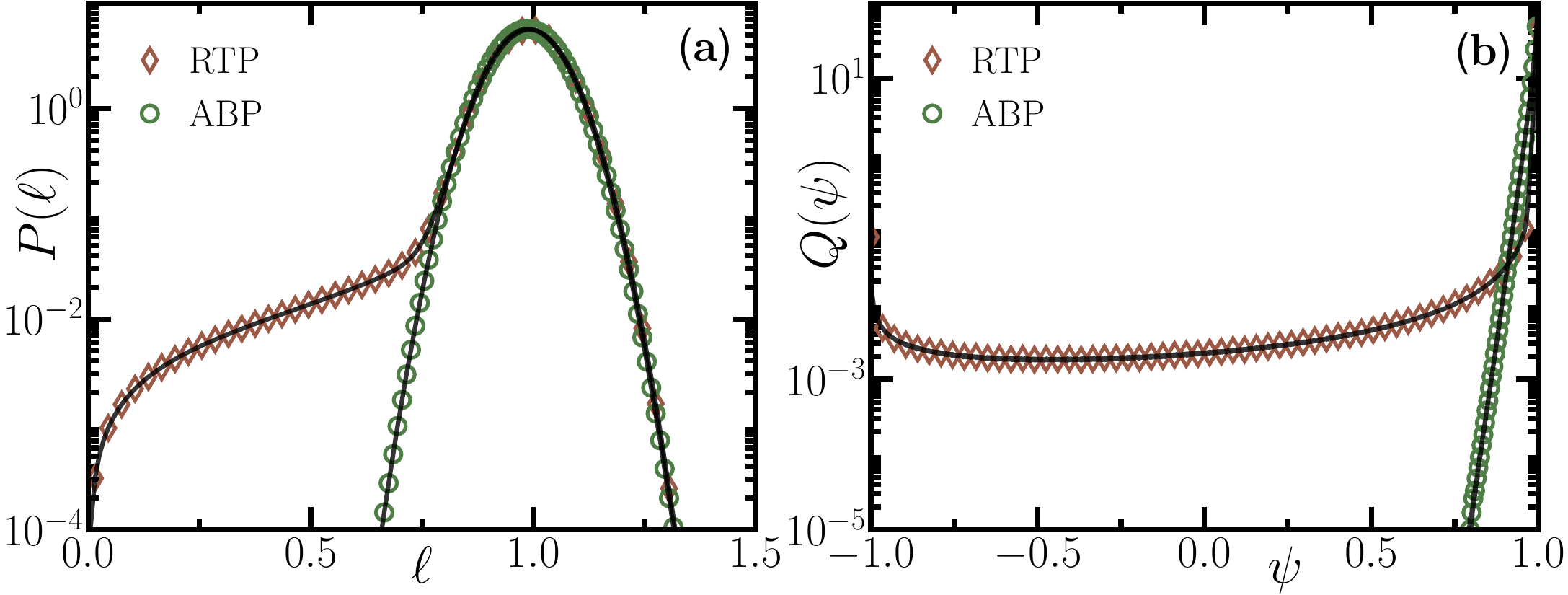}
    \caption{Comparison of RTP and ABP dynamics: Plots of (a)  Radial position distribution $P(\ell)$ and (b) radial polarization distribution $Q(\psi)$ for RTP and ABP with the same parameters. The symbols indicate data obtained from numerical simulations, whereas the black solid lines denote the corresponding theoretical predictions [see \eref{eq:Pl_RTP}, \eqref{eq:Pl_ABP}, \eqref{eq:Q_F_s} and \eqref{eq:Ppsi_ABP}]. Here we have used $N=100, k=0.01$ and $v_0=1$, so that $R_0=0.99$.}
    \label{fig:compare}
\end{figure}

\subsection{Active Brownian particles}\label{sec:Ppsi_ABP}

To derive the conditional distribution $P({\bm \ell}_a| \hat{\bm n})$ for the ABP dynamics, 
we start from the Langevin equation \eqref{eq:Le_active}. For $\alpha=0$, the orientation vector $\hat{\bm n}$ remains frozen and, for a given $\hat{\bm n}$, the position deterministically relaxes to  ${\bm \ell}_a^{(0)}=R_0\hat{\bm n}$. For non-zero but small $\alpha$, the orientation vector evolves in time, and we expect that, in the stationary state, \eref{eq:Le_active} would admit a solution of the form,
\begin{align}
    {\bm \ell}_a = R_0\hat{\bm n} + {\bm w},\label{eq:la_simple}
\end{align}
where, ${\bm w}$ denotes fluctuations around the deterministic solution.  Substituting the above equation in \eref{eq:Le_active}, we find that ${\bm w}$  must satisfy,
\begin{align}
    \dot{\bm w} = -\mu{\bm w} - R_0\dot{\hat{\bm n}}.\label{eq:Le_w}
\end{align}
It is now convenient to define the unit vector $\hat{\bm e}=(-\sin\theta, \cos\theta)$ perpendicular to the orientation vector $\hat{\bm n}$. Using the decomposition, ${\bm w}=w_{\shortparallel}\hat{\bm n}+w_{\perp}\hat{\bm e}$, it can be shown [see Appendix~\ref{ap:Ppsi_ABP}] that $w_{\shortparallel}$ is a damped process with no noise source. Therefore in the stationary limit $w_{\shortparallel}=0$ to $O(\sqrt{1/\tau})$. The perpendicular component on the other hand follows an OU process, which admits a Gaussian distribution with mean zero and variance $\alpha R_0^2$. Consequently, we have,
\begin{align}
    P({\bm w}) = \delta(w_{\shortparallel})\frac{1}{\sqrt{2\pi\alpha}R_0} e^{-w_{\perp}^2/(2\alpha R_0^2)}.\label{eq:Pw}
\end{align}
The above equation together with \eref{eq:la_simple} gives the conditional distribution,
\begin{align}
    P({\bm \ell}_a|\hat{\bm n}) = \delta(\ell_{a,\shortparallel}-R_0)\frac{1}{\sqrt{2\pi\alpha}R_0}e^{-\ell_{a,\perp}^2/(2\alpha R_0^2)}.\label{eq:P_lan_cond}
\end{align}
where we have used the decomposition ${\bm \ell}_a=\ell_{a,\shortparallel}\hat{\bm n}+\ell_{a,\perp}\hat{\bm e}$. Finally, substituting  \eref{eq:P_lan_cond} in \eref{eq:P_ln_cond} and carrying out the integral over ${\bm \ell}_a = (\ell_{a,\shortparallel},\ell_{a,\perp})$  yields the anisotropic Gaussian distribution,
\begin{align}
    P({\bm \ell}|\hat{\bm n}) = \frac{1}{2\pi\sigma_{\shortparallel}\sigma_{\perp}} \exp\Big[-\frac{(\ell_{\shortparallel}-R_0)^2}{2\sigma_{\shortparallel}^2}-\frac{\ell_{\perp}^2}{2\sigma_{\perp}^2}\Big], \label{eq:P_ln_gauss}
\end{align}
where, we have used ${\bm \ell}=\ell_{\shortparallel}\hat{\bm n}+\ell_\perp\hat{\bm e}$ and defined $\sigma^2_{\shortparallel}=\sigma^2/2$ and $\sigma^2_{\perp}=\sigma^2/2+\alpha R_0^2$. 
Using the above equation in \eref{eq:PPsi_formal} and evaluating the integral over ${\bm \ell}$ gives the radial polarization distribution for the ABP dynamics in the strongly active regime,
\begin{align}
    Q(\psi) = \frac{e^{-R_0^2/(2\sigma^2_{\shortparallel})}/A^2(\psi)}{\pi\sigma_{\shortparallel}\sigma_{\perp}\sqrt{1-\psi^2}}\Bigg[1+ \sqrt{\frac \pi 2}\frac{B(\psi)}{A(\psi)}~e^{\frac{B^2(\psi)}{2A^2(\psi)}}\text{erfc}\left(-\frac{B(\psi)}{\sqrt{2}A(\psi)}\right)\Bigg],\label{eq:Ppsi_ABP}
\end{align}
where, $\text{erfc}(x)$ denotes the complementary error function and we have defined,
\begin{align}
    A^2(\psi)=\frac{\psi^2}{\sigma_{\shortparallel}^2}+\frac{1-\psi^2}{\sigma^2_{\perp}},\quad B(\psi)=\frac{R_0\psi}{\sigma^2_{\shortparallel}}.   
\end{align}
Figure~\ref{fig:Ppsi}(b) shows plot of the radial polarization distribution $Q(\psi)$ obtained from numerical simulations along with the theoretical prediction~\eref{eq:Ppsi_ABP} for different values of $\alpha$. The excellent agreement between the two validates our theoretical prediction and the assumptions behind it. In the limit $\alpha\to0$, $\sigma_{\shortparallel}^2=\sigma_{\perp}^2=\sigma^2/2$, $A(\psi)=2/\sigma^2,B(\psi)=2R_0\psi/\sigma^2$, and the above expression simplifies to \eref{eq:PPsi_a0}.

\begin{figure}
    \centering
    \includegraphics[width=8.5cm]{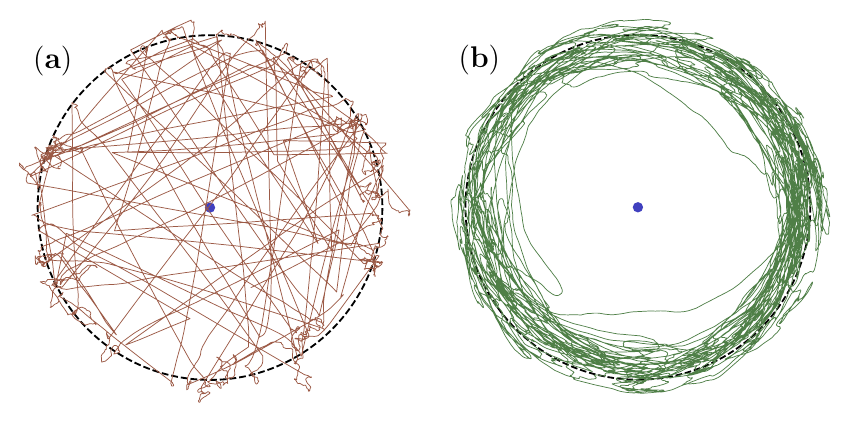}
    \caption{Typical trajectory of a single tagged particle from a collection of $N=100$ harmonically coupled (a) Run-and-Tumble particles and (b) active Brownian particles. The blue dot indicates the position of the centroid. In both panels, the black dashed lines indicate a circle of radius $R_0$. Here we have taken $\tau=10$, $k=0.01$ and $v_0=1$ so that $R_0=0.99$.}
    \label{fig:trajectory}
\end{figure}

\section{Discussion: ABP versus RTP}\label{sec:comp}

We now turn to a direct comparison of the spatial and orientational organization of ABPs and RTPs, focusing on how the differences arise from the propulsion mechanisms. As we have seen from the previous sections, these differences are 
manifest in the measures of both spatial and orientational organizations, especially in the strongly active regime. Although the mathematical results for the two systems have already been presented in the preceding sections, we include this comparative discussion to consolidate the main findings and provide a unified perspective.

Let us first focus on the spatial organization, characterized by the distribution $P(\ell)$ of the radial distance $\ell$ of a tagged particle from the centroid. Figure~\ref{fig:compare}(a) compares this distribution for RTP and ABP for the same of parameters, in the strongly active regime $\alpha \ll 1$. Clearly, the Gaussian-like peak, which arises from the leading order contribution in the $\alpha \to 0$ limit, remains same [see \eref{eq:P0_active}]. However, the first order correction $O(\alpha)$ are very different in the two cases, giving rise to the shoulder-like structure for the RTP dynamics. Physically, $\alpha \to 0$ corresponding to the infinite persistence time limit, that is when the orientations of the particles barely evolve with time. The correction to this limit comes from the scenario when the orientation of a particle changes with time, albeit with a large (but finite) persistence time. For an RTP this implies a random finite change in the orientation, while for an ABP the change is continuos. This leads to the trajectories of the particles being very different, which is illustrated in Fig.~\ref{fig:trajectory}. The large changes in orientation leads to the RTP wandering away from the circle more frequently, which, in turn, leads to an increased probability of its being near the origin.

To understand the difference in the radial distributions of ABP and RTP more comprehensively it is also useful to look at the corresponding moments [see Fig.~\ref{fig:l_moments}]. From the characteristic function $\tilde P(q)$ given in \eref{eq:Pq_1}, it is straightforward to relate 
moments of $\ell^2$ to that of $\ell_a$. In particular, we have,
\begin{align}
    \la \ell^2 \ra &= \sigma^2+\la \ell_a^2 \ra, ~~\text{and}, ~~
    \la \ell^4 \ra = 2\sigma^4+4\sigma^2\la \ell_a^2\ra+\la \ell_a^4 \ra,\label{eq:l24_def}
\end{align}
where $\la \ell_a^n \ra$ denotes the $n$-th moment of the active process \eref{eq:Le_active} and $\sigma^2$ is given in \eref{eq:G_OU}. 

As mentioned already, the two-point auto-correlation of the orientation vector is same for both ABP and dynamics [see \eref{eq:n_auto}]. This leads to $\la \ell_a^2 \ra = R_0^2/(\alpha +1)$ for both cases, which, in turn, leads to $\la \ell^2 \ra$ being identical for ABP and RTP [see Fig.~\ref{fig:l_moments}]. The difference becomes apparent from higher order moments. For example, from \eref{eq:P_RTP}, we have, for RTP,
\bea 
\la \ell_a^4 \ra = \frac{2 R_0^4}{(\alpha+1)(\alpha+2)}. \label{eq:l4_avg_RTP}
\eea
On the other hand, for the fourth moment of ABP is given by~\cite{Chaudhuri_2021},
\bea 
\la \ell_a^4 \ra = \frac{R_0^4(4\alpha+3)}{(\alpha+1)(2\alpha+1) (\alpha+3)}.\label{eq:l4_avg_ABP}
\eea
This leads to the sifnificant difference in fourth moment $\la \ell^4 \ra$ for RTP and ABP dynamics, as shown in Fig.~\ref{fig:l_moments}. 

As discussed in Sec.~\ref{sec:Ppsi}, the difference between ABP and RTP system becomes more apparent from fluctuations of the radial polarization $\psi$. This is illustrated in Fig.~\ref{fig:compare}(b) which compares $Q(\psi)$ for same parameter values for ABP and RTP dynamics in the strongly active regime $\alpha \ll 1$. The drastic difference between between the two cases also physically originates from the short-time dynamics of orientation vector being distinct in the two cases. 

The fixed orientations of the particles in the  $\alpha \to 0$ limit corresponds to $\psi=1$ for both cases. For ABP, in the small $\alpha >0$ regime, the orientation evolves slowly, giving rise to the strong peak at $\psi=1$. On the other hand, the random (possibly large) changes in the orientation for RTP leads to finite probabilities of the orientation even being anti-parallel to the radial vector. Consequently, $Q(\psi)$ acquires finite weight for all values of $\psi$. 

It should be mentioned here that the ring-like formation studied in this work is also expected to emerge for other active particle dynamics with constant propulsion speed like direction reversing active Brownian particles~\cite{santra2021DRABP}. On the other hand, for active Ornestein-Uhlenbeck particles, formation of such a structure is not possible due to the  the Gaussian nature of the noise.

\section{Conclusion}\label{sec:conclusion}
We study the collective dynamics of a system of $N$ active particles coupled via attractive harmonic potential and show that a ring-like structure emerges in the stationary state for both Run-and-Tumble and active Brownian particles. We study this emergent spatial structure by analytically characterising the fluctuations of the radial distance of a tagged particle from the centroid and its radial polarization. Despite the apparent universality of the emergent ring-like structure, we find that its fluctuation statistics are strongly model dependent in the highly active regime. We trace these differences to the distinct short-time persistence and reorientation dynamics of RTPs and ABPs, demonstrating that the microscopic propulsion mechanism leaves a clear and measurable imprint on the stationary collective state.

An obvious question that remains open is that how this emergent structures and the role of propulsion mechanisms change when the coupling changes, in particular what happens when the interaction is non-linear and/or short-ranged. Another intriguing question is what happens when we consider a system with dispersion in activity~\cite{sarkar2025emergent, coupled_ABP_long_2025,lauersdorf2025binary,dutta2026dispersion}. Finally, it would also be interesting to explore the emergent organisation by an interacting system inertia ABPs~\cite{patel2023IABP,patel2024exact} and RTPs~\cite{dutta2024harmonically,dutta2025inertial}.

\appendix

\renewcommand{\theequation}{\Alph{section}.\arabic{equation}}
\setcounter{equation}{0}

\section{Characteristic function of $P(\ell)$} \label{ap:Pq}

In this appendix we provide details of the computation leading to the general form of the  characteristic function of $P(\ell)$, given in \eref{eq:Pq_1} in the main text. Taking the Fourier transform of \eref{eq:dist_l} with respect to $q$,
we get,
\begin{align}
    \tilde{P}(q) = \intop_{0}^{\infty}d\ell_a P(\ell_a)e^{i q\,\ell_{a}^{2}}\intop_{0}^{\infty}d\ell_g P(\ell_g)e^{i q\,\ell_{g}^{2}}\intop_0^{2\pi}\frac{d\phi}{2\pi}e^{i(2q\ell_a\ell_g)\cos\phi}.\label{eq:Pq}
\end{align}
The integral over $\phi$ can be immediately carried out, giving,
\begin{align}
    \intop_0^{2\pi}\frac{d\phi}{2\pi}e^{i(2q\ell_a\ell_g)\cos\phi}=J_{0}(2q\ell_a\ell_g),
\end{align}
where, $J_0(z)$ is the zeroth order Bessel function. Substituting the above equation in \eref{eq:Pq}, we get,
\begin{align}
    \tilde{P}(q)=\intop_0^{\infty} d\ell_a e^{iq \ell^2_a} {\cal P}(\ell_a) \intop_0^\infty d\ell_g e^{iq \ell^2_g}  G(\ell_g) J_{0}(2q\ell_a\ell_g).
\end{align}
The integral over $\ell_g$ can be exactly performed using the Gaussian distribution $G(\ell_g)$ given in \eref{eq:G_OU}, which leads to \eref{eq:Pq_1} quoted in the main text.

\section[Radial distribution for RTP]{Computation of the radial distribution for the RTP dynamics} \label{ap:Pl_RTP}

\setcounter{equation}{0}

In this appendix, we derive the expression for the radial distribution $P(\ell)$ of the RTPs from the characteristic function \eqref{eq:RTP_char}, as well as its asymptotic forms in the strongly active and passive regimes. We start by substituting the integral representation of the confluent hypergeometric function,
\begin{align}
    _1F_1(1,1+\alpha,x)=\alpha\intop_0^1dt\,e^{x\,t}(1-t)^{\alpha-1},
\end{align}
in \eref{eq:RTP_char}. Subsequently taking the inverse Fourier transform of $\tilde P(q)$ in \eref{eq:RTP_char}  gives,
\begin{align}
    \mathbb{P}(z) = \frac{i\alpha}{2\pi}\intop_0^{1}dt(1-t)^{\alpha-1}\intop_{-\infty}^{\infty}\frac{dq}{i+q\sigma^2}\exp\left[-q\left(iz+\frac{R_0^2 t}{i+q\sigma^2}\right)\right].
\end{align}
Evaluating the integral over $q$ yields,
\begin{align}
    \mathbb{P}(z) = \frac{\alpha}{\sigma^2}e^{-z/\sigma^2}\intop_0^1 dt(1-t)^{\alpha-1}e^{-R_0^2t/\sigma^2}I_0\left(\frac{2R_0\sqrt{zt}}{\sigma^2}\right).
\end{align}
Finally substituting the above expression in \eref{eq:Pzl}, we obtain \eref{eq:Pl_RTP} quoted in the main text.

We now discuss the behaviour of the radial distribution in the different limits of  activity. We first consider the strongly active regime $\tau\gg1/\mu$ , i.e., $\alpha \ll 1$. To this end, we expand the right hand side of \eref{eq:PlRTP_series} upto linear order in $\alpha$ around $\alpha=0$, using, 
\begin{align}
    \alpha\Gamma(\alpha)&\simeq (1-E_\gamma\, \alpha),\\ 
    {}_{1}{F}_1(k+1,k+1+\alpha,-z)&\simeq e^{-w} +\alpha \left .\partial_{\alpha}\big[{}_1\tilde{F}_{1}(k+1,k+1+\alpha,-w)\big]\right|_{\alpha=0}.
\end{align}
This leads to \eref{eq:P1_RTP} in the main text. \\

\noindent{\bf Small $\ell$ behaviour of $P(\ell)$:} From \eref{eq:Pl_RTP}, we have, near $\ell=0$, 
\begin{align}
    P(\ell)\approx\frac{2\ell}{\sigma^2}e^{-\ell^2/\sigma^2}{}_1\tilde{F}_{1}(1;\alpha+1;-\frac{R_0^2}{\sigma^2}).
\end{align}
In the strongly active regime, $\alpha\ll1$, and $R_0^2/\sigma^2\approx N\gg1$, and we have to order $O(1/\tau)$,
\begin{align}
    P(\ell)\approx\frac{2\ell}{\sigma^2}e^{-\ell^2/\sigma^2}\Big[e^{-R_0^2/\sigma^2}+\frac{\alpha R_0^2}{\sigma^2}\Big],
\end{align}
which leads to \eref{eq:P0_sml_RTP} and \eref{eq:P1_sml_RTP} in the main text. \\

\noindent{\bf Strongly passive regime:} In the strongly passive regime $\tau\ll\mu$, i.e., $\alpha\gg1$, and we can substitute the approximation $(1-t)^{\alpha-1}\approx e^{-\alpha t}$ in \eref{eq:Pl_RTP}, which now becomes,
\begin{align}
    P(\ell)\approx\frac{2\alpha\ell}{\sigma^2}e^{-\ell^2/\sigma^2}\intop_0^{1}dt\,e^{-(\alpha-R_0^2/\sigma^2)t}I_0\Big(\frac{2R_0\ell}{\sigma^2}\sqrt{t}\Big).
\end{align}
Using the substitution $t=u/\alpha$, the integral can be approximated as,
\begin{align}
    P(\ell)\approx\frac{2\ell}{\sigma^2}e^{-\ell^2/\sigma^2}\intop_0^{\infty}du\,\exp\Big[-\Big(1+\frac{R_0^2}{\alpha \sigma^2}\Big)u\Big]I_0\Big(\frac{2R_0\ell}{\sigma^2}\sqrt{\frac{u}{\alpha}}\Big).
\end{align}
The above integral can be carried out exactly which leads to the Rayleigh distribution \eref{eq:rayleigh} in the passive regime.

\section[Radial distribution for ABP]{Computation of the radial distribution for the
ABP dynamics }\label{ap:Pl_ABP}
\setcounter{equation}{0}

Here we provide the details of computation leading to the results reported in Sec.~\ref{sec:Pl_ABP}. We start with the derivation of \eref{eq:ABP_char}. Substituting \eref{eq:P_ABP} in \eref{eq:Pq_1}, we get,
\begin{align}
    \tilde{P}(q)=\frac{i}{i+q\sigma^2}\frac{1}{D_t}\sqrt{\frac{2\pi \mu}{\tau}}\sum_{n=0}^{\infty}\Big(\frac{b^2v_0^2\tau}{D_t}\Big)^n C_{n,0}~f_{n}(q),\label{eq:Pq_2}
\end{align}
where, for notational simplicity, we have defined,
\begin{align}
    f_{n}(q) = \intop_{0}^{\infty}d\ell_\text{A}\,\ell_\text{A} e^{-\mu \ell_\text{A}^2 s(q)/(2 D_t)}L_{n}\Big(\frac{\mu\ell_\text{A}}{2 D_t}\Big),
\end{align}
with, 
\begin{align}
    s(q)=1+\frac{2 q D_t}{\mu(i+q\sigma^2)}.
\end{align}
Using the change of variable $x=\mu\ell_\text{A}^2/(2D_t)$, the integral $f_n(q)$ can be recast in the form,
\begin{align}
    f_n(q) = \frac{D_t}{\mu}\intop_0^{\infty} dx\,e^{-x s(q)}L_{n}(x).
\end{align}
It is now straightforward to evaluate the integral using the standard identity\cite[\href{https://dlmf.nist.gov/18.17.E40}{(18.17.40)}]{NIST:DLMF}, yielding,
\begin{align}
    f_{n}(q) = \frac{D_t}{\mu}\frac{(s(q)-1)^{n}}{s(q)^{n+1}}. 
\end{align}
Substituting the above form in \eqref{eq:Pq_2}, we  obtain 
\eref{eq:ABP_char} for 
the characteristic function for ABP.

To compute the corresponding inverse Fourier transform, it is convenient to define
\begin{align}
    g_n(z)=\frac{1}{2\pi}\intop_{-\infty}^{\infty}dq\frac{e^{-iqz}q^n}{(i+q\sigma^2)^{n+1}}. \label{eq:gn_z}
\end{align}
To evaluate the $q$-integral, we first note that \eref{eq:gn_z} can be recast in the form,
\begin{align}
    g_{n}(z) = \frac{(i\partial_z)^n}{\sigma^{2n+2}}\frac{1}{2\pi}\intop_{-\infty}^{\infty}dq~\frac{e^{-iqz}}{(q+i/\sigma^2)^{n+1}}
\end{align}
The integral in the above equation can be computed explicitly using a closed semi-circular contour in the lower half plane, which leads to,
\begin{align}
    g_{n}(z) = -\frac{i}{\sigma^{2n+2}\,n!}\partial_z^{n}\big[z^ne^{-z^2/\sigma^2}\big]=-\frac{i\,e^{-z/\sigma^2}}{\sigma^{2n+2}}L_{n}\Big(\frac{z}{\sigma^2}\Big)\label{eq:contour}
\end{align}
Note that, in the second step, we have used the Rodrigues' formula\cite[\href{https://dlmf.nist.gov/18.9.E24}{(18.9.24)}]{NIST:DLMF} for Laguerre polynomials.Using \eref{eq:contour} along with \eqref{eq:Pz_1} we explicitly obtain the position distribution \eqref{eq:Pl_ABP} in the main text. \\

\noindent{\bf Strongly active regime:} We now compute the asymptotic form of the radial distribution in the  strongly active limit ($\alpha \ll 1$). To this end, we approximate the coefficients $C_{n,l}$ to the leading orders in $\alpha=1/(\tau\mu)$ as,
\begin{align}
    C_{n,l}\approx C_{n,l}^{(0)}(1+\alpha a_{n,l}), \label{eq:C_approx}
\end{align}
where, $C_{n,l}^{(0)}$ are defined in \eref{eq:C_0}. Using \erefs{eq:C_approx} and \eqref{eq:C_0} along with \eref{eq:recur}, we can obtain a recursion relation for the corrections $a_{n,l}$. For this it is first convenient to compute the ratios $C_{n,l-1}^{(0)}/C_{n,l}^{(0)}$ and $C_{n-1,l+1}^{(0)}/C_{n,l}^{(0)}$ using \eref{eq:C_0} which are given by,
\begin{align}
    \frac{C_{n,l-1}^{(0)}}{C_{n,l}^{(0)}} = \sqrt{2(n+l)/\alpha},\quad
    \frac{C_{n-1,l+1}^{(0)}}{C_{n,l}^{(0)}} = -\sqrt{2n/\alpha}.\label{eq:Cr}
\end{align}
Substituting \eref{eq:C_approx} in \eref{eq:C_recur_1} and using \eqref{eq:Cr}, we obtain the leading order corrections in $\alpha$,
\begin{align}
    C_{n,l}\approx C_{n,l}^{(0)}\Big(1+\frac{\alpha}{2n+l}\Big[(n+l)a_{n,l-1}+n a_{n-1,l+1}-l^2\Big]\Big).
\end{align}
Comparing the above equation with \eref{eq:C_approx}, we obtain the recursion relation for the coefficients $a_{n,l}$,
\begin{align}
    a_{n,l}=\frac{1}{2n+l}\Big[(n+l)a_{n,l-1}-n a_{n-1,l+1}-l^2\Big].\label{eq:a_rec}
\end{align}
The boundary conditions for $a_{n,l}$ can be obtained by setting $n=0$ in \eref{eq:C_recur_1}, and solving the corresponding recursion relation for $C_{0,l}$. Keeping the terms to order $O(\alpha)$,
\begin{align}
    C_{0,l}\approx C_{0,l}^{(0)}\Big[1-\frac{\alpha}{2}l(l+1)\Big].
\end{align}
Comparing with \eref{eq:C_approx}, gives the first boundary condition,
\begin{align}
    a_{0,l}=-\frac{l(l+1)}{2}.\label{eq:a_bc_1}
\end{align}
Similarly, the other boundary condition can be obtained by substituting \eref{eq:C_approx} in \eref{eq:C_recur_2}, yielding,
\begin{align}
    a_{n,0}=a_{n-1,1}.\label{eq:a_bc_2}
\end{align}
Equations~\eqref{eq:a_rec} together with \eqref{eq:a_bc_1} and \eqref{eq:a_bc_2} give the coefficients explicitly,
\begin{align}
    a_{n,l} = -n -\frac{l(l+1)}{2}.
\end{align}
This, in turn, leads to \eref{eq:C_approx} in the main text.

To get the radial distribution in the strongly active regime, we substitute the above expression for the coefficients in \eref{eq:Pl_ABP}, yielding,
\begin{align}
    P(\ell) = \frac{2\ell}{\sigma^2}e^{-\ell^2/\sigma^2}\sum_{n=0}^{\infty}\frac{(-R_0^2/\sigma^2)^n}{n!}L_{n}\Big(\frac{\ell^2}{\sigma^2}\Big)(1-\alpha n).\label{eq:Pl_ABP_2}
\end{align}
Substituting the explicit form for $L_{n}(x)$,  $P(\ell)$ can be expressed as,
\begin{align}
    P(\ell)=\frac{2\ell}{\sigma^2}e^{-\ell^2/\sigma^2}\sum_{n=0}^{\infty}\sum_{m=0}^{n}\frac{(-\ell^2/\sigma^2)^m(-R_0^2/\sigma^2)^n}{(m!)^2(n-m)!}(1-\alpha n).
\end{align}
This double sum can be evaluated explicitly and results in \eref{eq:P1_active_ABP} in the main text.\\

\noindent{\bf Strongly passive regime:} In the passive regime $\alpha\gg1$, the coefficients $C_{n,0}$ to the order $O(1/\alpha)$ can be approximated as\cite{malakar2020steady},
\begin{align}
    C_{n,0}=\sqrt{\frac{1}{2\pi\alpha}}\Big(-\frac{1}{2}\Big)^{n}\Big[1-\frac{1}{8\alpha}(7n^2+n)\Big].
\end{align}
Substituting the above approximation in \eref{eq:Pl_ABP} gives,
\begin{align}
    P(\ell)\approx \frac{2\ell}{\sigma^2}e^{-\ell^2/\sigma^2}\sum_{n=0}^{\infty}\left(-\frac{\mu\tau R_0^2}{\sigma^2}\right)^{n}L_{n}\Big(\frac{\ell^2}{\sigma^2}\Big)\Big[1-\frac{1}{8\alpha}(7n^2+n)\Big].\label{eq:Pl_passive_sum_ABP}
\end{align}
This sum can be evaluated exactly using the generating function of Laguerre polynomials,
\begin{align}
   \frac{e^{-a x/(1-a)}}{(1-a)}=\sum_{n=0}^\infty a^n L_n(x).
\end{align}
Simplifying the resulting expression in the large $\alpha \gg 1$ regime we get the correction to the Rayleigh distribution \eqref{eq:rayleigh} in the passive regime, which is quoted in \eref{eq:rayleigh_corr} in the main text.

\section{Radial polarization distribution of Run-and-Tumble particles}\label{ap:Ppsi_RTP}
\setcounter{equation}{0}

In this appendix we simplify $F(\psi,s)$ and explicitly compute $F(\psi,0)$ given in Sec.~\ref{sec:Ppsi_RTP}. We start by recasting the integral in \eref{eq:Fs} in terms of the angle $\chi$ between $\hat{\bm n}$ and ${\bm \ell}$ and the angle $\vartheta$ between $\hat{\bm n}'$ and  ${\bm \ell}$,
\begin{align}
    F(\psi,s)=& \frac{1}{2\pi^2\sigma^2}\intop_0^{2\pi}d\chi\intop_0^{2\pi}d\vartheta\, e^{-R_0^2 \left[(1-s)^2+s^2+2s(1-s)\cos(\vartheta-\chi)\right]/\sigma^2} \cr 
   \times &\intop_0^\infty d\ell\,\ell \exp{\left[2\ell R_0((1-s)\cos\chi+s\cos\vartheta)-\ell^2\right]} \delta(\psi-\cos\chi).
\end{align}
Evaluating the integral over $\chi$ and $\ell$, yields,
\begin{align} 
    F(\psi,s) =\frac 1{\sqrt{1- \psi^2}}\intop_0^{2 \pi} \frac{d \vartheta}{2 \pi^2} e^{-\frac{R_0^2 \lambda^2}{\sigma^2}} \left[1+ \sqrt{\pi} \frac{R_0 \epsilon}{\sigma} e^{\frac{R_0^2\epsilon^2}{\sigma^2}} \text{erfc}\left(-\frac{\epsilon R_0}{\sigma}\right)\right],
\end{align}
where we have used the definitions,
\begin{align}
    \epsilon &=(1-s)\psi+s\cos \vartheta, ~\text{and},\\
    \lambda^2 &=(1-s)^2+s^2+2s(1-s)\cos(\vartheta-cos^{-1}\psi).
\end{align}
Substituting $s=0$ in the above expression, so that $\lambda^2=1$ and $\epsilon=\psi$, we get \eref{eq:PPsi_a0} given in the main text. The linear order corrections in $\alpha$ can be computed by numerically evaluating the $\vartheta$ itegral and using it in \eref{eq:Q_F_s}.

\section{Radial polarization distribution of active Brownian particles}\label{ap:Ppsi_ABP}
\setcounter{equation}{0}

In this appendix we provide the details of the computation leading to the results in Sec.~\ref{sec:Ppsi_ABP}. We start from the time-evolution of the orientation vector $\hat{\bm n}$ and its perpendicular $\hat {\bm e}$. From the Brownian dynamics of the orientation angle $\theta$ [see \eref{eq:tht_evol}], we get, 
\begin{align}
    \dot{\hat{\bm n}}=\dot{\theta}~\hat{\bm e},\quad\text{and},\quad\dot{\hat{\bm e}}=\dot{\theta}~\hat{\bm n}.
\end{align}
Using the decomposition, ${\bm w}=w_{\shortparallel}\hat{\bm n}+w_{\perp}\hat{\bm e}$ introduced in the main text, \eref{eq:Le_w} can be recast component wise,
\begin{align}
    \dot{w}_{\shortparallel}&=-\mu w_{\shortparallel}+w_{\perp}\dot{\theta},\\
    \dot{w}_{\perp}&=-\mu w_{\perp}-(R_0+w_{\shortparallel})\dot{\theta}.
\end{align}
Now, $\dot{\theta}=O(\sqrt{1/\tau})$ and ${\bm w}$, which in turn is driven by $\dot{\theta}$ is also $O(\sqrt{1/\tau})$. Therefore, to order $O(\sqrt{1/\tau})$ the above equations become,
\begin{align}
    \dot{w}_{\shortparallel}\simeq-\mu w_{\shortparallel},~~~\dot{w}_{\perp}\simeq-\mu w_{\perp}-R_0\sqrt{2/\tau}~\eta(t).
\end{align}
This leads to \eref{eq:Pw} and subsequently \eref{eq:P_ln_gauss} in the main text. 

At the next step, to compute $Q(\psi)$, we substitute \eref{eq:P_ln_gauss} into \eref{eq:PPsi_formal}. Performing the $\theta$ integral gives,
\begin{align}
    Q(\psi)=\frac{1}{\pi\sigma_{\shortparallel}\sigma_{\perp}\sqrt{1-\psi^2}}\intop_0^{\infty}d\ell\,\ell\exp\Big[-\frac{(\ell\psi-R_0)^2}{2\sigma^2_{\shortparallel}}-\frac{\ell^2(1-\psi^2)}{2\sigma^2_{\perp}}\Big].
\end{align}
Finally evaluating the integral in $\ell$ leads to an explicit form of $Q(\psi)$ given in \eref{eq:Ppsi_ABP} in the main text. 

\bibliography{ring.bib} 
\bibliographystyle{rsc} 

\end{document}